\documentclass[12pt]{article}
\usepackage{amssymb,amsmath,epsfig}
\allowdisplaybreaks
\begin{document}
\title{\bf Inhomogeneous Perturbations and Stability Analysis of the Einstein Static Universe in
$f(R,T)$ Gravity}

\author{M. Sharif \thanks{msharif.math@pu.edu.pk} and Arfa Waseem
\thanks{arfawaseem.pu@gmail.com}\\
Department of Mathematics, University of the Punjab,\\
Quaid-e-Azam Campus, Lahore-54590, Pakistan.}
\date{}

\maketitle

\begin{abstract}
The purpose of this paper is to analyze the existence of static
stable Einstein universe using inhomogeneous linear perturbations in
the context of $f(R,T)$ gravity ($R$ and $T$ denote the scalar
curvature and trace of the stress-energy tensor, respectively). The
static and perturbed field equations are constructed for perfect
fluid parameterized by linear equation of state parameter. We obtain
solutions manifesting the Einstein static state by considering
peculiar $f(R,T)$ forms for vanishing and non-vanishing conservation
of the stress-energy tensor. It is observed that stable static
Einstein regions exist for both closed as well as open FLRW universe
models for an appropriate choice of parameters. We conclude that
this theory is efficient for presenting such cosmological solutions
leading to emergent universe scenario.
\end{abstract}

\section{Introduction}

Modified gravitational theories have attracted many researchers to
discuss the accelerated expanding universe. The direct modification
of general relativity (GR) is the $f(R)$ theory which is derived by
introducing a general function $f(R)$ in place of scalar curvature
$(R)$ in Einstein theory (Capozziello 2002; Nojiri and Odintsov
2003). One of the most stimulating characteristics of extended
theories is the inclusion of coupling between gravitational and
matter entities that has instigated several researchers to unveil
the hidden mysteries of dark components. Harko et al. (2011)
established such type of interaction in $f(R,T)$ theory. This
modified theory can be regarded as an extended form of $f(R)$
theory. The motivation of introducing trace of the energy-momentum
tensor (EMT) may originate from the results of some unknown
gravitational interactions or the effects of some exotic fluid. It
is predicted that such coupling provides the non-vanishing
conservation of EMT. Consequently, an additional force arises due to
which massive test particles follow the non-geodesic path while dust
particles chase the geodesic lines. The $f(R,T)$ theory has
extensively been studied for different tasks such as thermodynamics
(Jamil et al. 2012; Sharif and Zubair 2012, 2013a), energy
conditions (Alvarenga et al. 2013; Sharif et al. 2013; Sharif and
Zubair 2013b), cosmological solutions (Shabani and Farhoudi 2013;
Sharif and Zubair 2014a, b; Moraes 2015), dynamical instability
(Noureen and Zubair 2015; Sharif and Waseem 2018a) and astrophysical
scenarios (Sharif and Siddiqa 2018; Sharif and Waseem 2018b, 2019a,
b; Deb et al. 2019; Maurya et al. 2019).

The big-bang singularity is another well-known issue in modern
cosmology. In order to resolve this singularity issue, various
speculations have been suggested to construct non-singular or past
eternal cosmological models. The emergent universe conjecture has
been developed in the background of GR (Gasperini and Veneziano
2003; Khoury et al. 2004) which says that the universe remains in an
Einstein static state and then emerges into inflationary phase of
cosmos (Ellis and Maartens 2004; Ellis et al. 2004). The successful
emergent universe conjecture also demands the presence of stable
Einstein universe (EU) with respect to all types of perturbations.
Einstein universe is characterized by
Friedmann-Lema\^{i}tre-Robertson-Walker (FLRW) spacetime with
perfect fluid. Initially, this model was favorite to interpret the
static universe but later, it was observed that for homogeneous and
isotropic perturbations, the EU exhibits unstable regions around
equilibrium state (Eddington 1930). It was also determined that EU
always remains neutrally stable for inhomogeneous vector/tensor
perturbations as long as the speed of sound fulfills the inequality
$c_{s}^{2}>\frac{1}{5}$ and unstable otherwise (Harrison 1967;
Barrow et al. 2003). Moreover, Barrow and Yamamoto (2012) analyzed
the stability of EU with homogeneous perturbations for different
kinds of matter fields and found unstable solutions.

Despite the fact that the basic component for developing emergent
scenario is the Einstein static solution, the initial model does not
prove successful to resolve the singularity issue, since scalar
homogenous perturbations break the stability of primal cosmic static
state in GR. Therefore, it was suggested to study Einstein static
cosmos beyond the Einstein gravity. In this respect, the stability
of EU has been investigated in several cosmological perspectives
such as brane-world gravity (Gergely and Maartens 2002), loop
quantum cosmology (Mulryne et al. 2005), Einstein Cartan scenario
(Atazadeh 2014) and scalar fluid theories (B\"{o}hmer et al. 2015).
The gravitational modified theories have become significant
approaches to derive stable Einstein solutions. B\"{o}hmer et al.
(2007) demonstrated the existence of stable EU solutions for
particular $f(R)$ functions by adopting homogeneous scalar
perturbations. They obtained stable EU by adding cosmological
constant and described the comparison with GR.

Goswami et al. (2008) discussed the stable EU modes in $f(R)$
background and observed the appearance of static Einstein solutions
only for $c_{s}>\sqrt{0.21}$ which is very close to the value
determined in GR. B\"{o}hmer and Lobo (2009) examined the same
scenario using linear homogeneous perturbations in $f(\mathcal{G})$
framework and constructed stable static solutions corresponding to
distinct values of equation of state (EoS) parameter. For generic
$f(R)$ models, Seahra and B\"{o}hmer (2009) inspected the stability
of EU with barotropic EoS and displayed that the unstable regions
are obtained for inhomogeneous perturbations. Li et al. (2013)
developed stable EU regions with respect to open and closed FLRW
models by applying homogeneous perturbations in generalized
teleparallel gravity. Huang et al. (2014) investigated the same
scenario using all kinds of perturbations in Jordan Brans-Dicke
theory. They also observed static EU by implementing perturbations
on matter variables in $f(\mathcal{G})$ gravity for closed cosmic
model (Huang et al. 2015).

In the context of curvature-matter coupled gravity, the conjecture
of emergent universe has gathered the attention of many researchers.
Shabani and Ziaie (2017) explored stable modes of EU in $f(R,T)$
background using perturbation technique as well as phase space
analysis. They obtained EU solutions corresponding to three
particular $f(R,T)$ models and analyzed their stability through
graphical analysis. On the same ground, Sharif and his collaborators
(2017; 2018; 2018c, d; 2019) obtained the stable EU regions by
considering homogeneous, inhomogeneous and anisotropic perturbations
in the framework of minimal as well as non-minimal coupled theories.
They also observed their solutions graphically and presented a
detail comparison with the existing literature.

This paper demonstrates the stability analysis of Einstein static
cosmos by employing scalar inhomogeneous perturbations in $f(R,T)$
framework. This study would be useful to analyze the role of
inhomogeneous perturbations as well as curvature-matter coupling on
the stable eras of EU. The next section manifests the formulation of
$f(R,T)$ field equations with respect to Einstein static state.
Section \textbf{3} provides the description about inhomogeneous
linear perturbations while section \textbf{4} deals with the
stability analysis of Einstein static solutions for conservation as
well as non-conservation of EMT corresponding to some particular
$f(R,T)$ models. The last section presents the summary of our work.

\section{Einstein Static Universe in $f(R,T)$ Scenario}

The $f(R,T)$ gravity with Lagrangian density of matter
($\mathcal{L}_{m}$) is characterized by the action (Harko et al.
2011)
\begin{equation}\label{1}
\mathcal{A}=\int\left(\frac{ f(R,T)}{2\kappa^{2}}+
\mathcal{L}_{m}\right)\sqrt{-g}d^4x,
\end{equation}
where $\kappa^{2}=1$ stands for coupling constant and $g$ indicates
determinant of the metric tensor ($g_{\gamma\eta}$). The EMT
corresponding to $\mathcal{L}_{m}$ is expressed as (Landau and
Lifshitz 1971)
\begin{equation}\label{2}
T^{\gamma\eta}=\frac{2}{\sqrt{-g}}\frac{\delta(\mathcal{L}_m
\sqrt{-g})}{\delta g_{\gamma\eta}}=g^{\gamma\eta}
\mathcal{L}_{m}+\frac{2\delta\mathcal{L}_{m}}{\delta
g_{\gamma\eta}}.
\end{equation}
The $f(R,T)$ field equations can be evaluated through varying the
action (\ref{1}) with respect to $g_{\gamma\eta}$ and are
represented by
\begin{eqnarray}\nonumber
R_{\gamma\eta}f_{R}(R,T)&-&\frac{1}{2}g_{\gamma\eta}f(R,T)
-(\nabla_{\gamma}\nabla_{\eta}-g_{\gamma\eta}\Box)f_{R}(R,T)
\\\label{3}&=&T_{\gamma\eta}-(\Theta_{\gamma\eta}
+T_{\gamma\eta})f_{T}(R,T),
\end{eqnarray}
where $f_{R}(R,T)$ and $f_{T}(R,T)$ depict differentiation of
generic function corresponding to $R$ and $T$, respectively,
$\Box=g^{\gamma\eta}\nabla_{\gamma}\nabla_{\eta}$, $\nabla_{\gamma}$
acts as the covariant derivative and $\Theta_{\gamma\eta}$ is
defined as
\begin{equation}\label{4}
\Theta_{\gamma\eta}=g^{\mu\nu}\frac{\delta T_{\mu\nu}}{\delta
g^{\gamma\eta}}= g_{\gamma\eta}\mathcal{L}_{m}-2T_{\gamma\eta}-
2g^{\mu\nu}\frac{\partial^{2}\mathcal{L}_{m}}{\partial
g^{\gamma\eta}\partial g^{\mu\nu}}.
\end{equation}
The covariant divergence of Eq.(\ref{3}) leads to
\begin{equation}\label{5}
\nabla^{\gamma}T_{\gamma\eta}=\frac{f_{T}}{1-f_{T}}
\left[(T_{\gamma\eta}+\Theta_{\gamma\eta})\nabla^{\gamma}(\ln
f_{T})-\frac{g_{\gamma\eta}}{2}\nabla^{\gamma}T+\nabla^{\gamma}
\Theta_{\gamma\eta}\right].
\end{equation}
We consider that the universe is comprised of perfect fluid given by
\begin{equation}\label{6}
T_{\gamma\eta}=(\rho+p)U_{\gamma}U_{\eta}- pg_{\gamma\eta},
\end{equation}
where $\rho$ denotes the matter density, isotropic pressure is
depicted by $p$ and $U_{\gamma}$ reveals the four velocity in
comoving frame. For perfect fluid configuration, we assume
$\mathcal{L}_{m}=-p$ which displays that matter Lagrangian depends
only on $g_{\gamma\eta}$ and not on its derivatives (Landau and
Lifshitz 1971). Hence, $\Theta_{\gamma\eta}=-2T_{\gamma\eta}-p
g_{\gamma\eta}$.

The emergent universe conjecture yields a viable alternative to the
initial singularity only with spatially non-flat FLRW spacetime
whose line element is expressed by (Huang et al. 2015)
\begin{equation}\label{7}
ds^{2}=a^{2}(\tau)\left[d\tau^{2}-\left(\frac{1}{1-\mathcal{K}
\chi^{2}}d\chi^{2}+\chi^{2}(d\theta^{2}+\sin^{2}\theta
d\phi^{2})\right)\right],
\end{equation}
where $a(\tau)$ manifests the conformal scale factor which is
determined by the conformal time $(\tau)$ and expresses the
connection as $a(\tau)d\tau=dt$ whereas $\mathcal{K}$ denotes the
parameter of spatial curvature which provides open, closed and flat
cosmic models for $\mathcal{K}=-1,1$ and $0$, respectively. The
scalar curvature and trace of EMT become
\begin{equation}\nonumber
R=-6\left(\frac{a\mathcal{K}+\ddot{a}}{a^{3}}\right),\quad
T=\rho-3p,
\end{equation}
where dot reveals differentiation associated with conformal time.
The corresponding field equations for the line element (\ref{7})
give
\begin{eqnarray}\nonumber
3\Big[\Big(\frac{\dot{a}}{a}\Big)^{2}+\mathcal{K}\Big]&=&\frac{1}
{f_{R}}\Big[\rho a^{2}+\frac{a^{2}}{2}f(R,T)+a^{2}(p+\rho)f_{T}+3
\Big(\mathcal{K}+\frac{\ddot{a}}{a}\Big)\Big.\\\label{8}&\times&\Big.
f_{R}-3\frac{\dot{a}}{a}\partial_{t}f_{R}\Big],
\\\nonumber\Big(\frac{\dot{a}}{a}\Big)^{2}-\frac{2\ddot{a}}{a}
-\mathcal{K}&=&\frac{1}{f_{R}}\Big[a^{2}p-\frac{a^{2}}{2}f(R,T)
-3\Big(\mathcal{K}+\frac{\ddot{a}}{a}\Big)f_{R}+\frac{\dot{a}}{a}
\partial_{t}f_{R}\Big.\\\label{9}&+&\Big.\partial_{tt}f_{R}\Big].
\end{eqnarray}

In the past decades, the question about the beginning and origin of
cosmos has provided fascinating results depending on the
observations of GR as well as modern cosmology. In accordance with
the fundamental physical perceptions on cosmic matter configuration,
GR equations signify that the current expanding cosmos must be
anticipated by a singularity, where the physical parameters such as
spacetime curvature and density diverge. To resolve this problem,
enormous research has been accomplished to assemble different
singularity free cosmological scenarios. In this respect, the
emergent universe conjecture has gained substantial importance to
solve the issue of primordial singularity (Gasperini and Veneziano
2003; Khoury et al. 2004). According to this conjecture, the
universe initiates asymptotically from Einstein static state and
then it moves into an expanding state that yields inflationary
scenario. The emergent universe model has interesting
characteristics like there is no primordial singularity, the
universe is eternal and static cosmic behavior in infinite past
$(t\rightarrow-\infty)$. Thus, the aim of this speculation is to
analyze the presence of stable Einstein solutions. For static EU
characterized by FLRW universe model, we consider $a(\tau)=a_{0}=$
constant and the associated forms of $R$ and $T$ provide
\begin{equation}\label{10}
R(a_{0})=R_{0}=-\frac{6\mathcal{K}}{a_{0}^{2}}, \quad
T_{0}=\rho_{0}-3p_{0},
\end{equation}
where $\rho_{0}$ and $p_{0}$ exhibit the static forms of matter
density and pressure, respectively. Equations (\ref{8}) and
(\ref{9}) turn into
\begin{eqnarray}\label{11}
3\mathcal{K}&=&\frac{1}{f_{R}}\left(a_{0}^{2}
\rho_{0}+\frac{a_{0}^{2}} {2}f(R_{0},T_{0})
+a_{0}^{2}(p_{0}+\rho_{0})f_{T}+3\mathcal{K}f_{R}\right),\\\label{12}
-\mathcal{K}&=&\frac{1}{f_{R}}\left(a_{0}^{2}p_{0}
-\frac{a_{0}^{2}}{2}f(R_{0},T_{0})-3\mathcal{K}f_{R}\right).
\end{eqnarray}
It is worth mentioning here that EU presents a rotation, expansion
as well as shear free cosmos. Using linear EoS $p=\omega \rho$ with
$\omega$ being an EoS parameter, addition of Eqs.(\ref{11}) and
(\ref{12}) leads to
\begin{equation}\label{13}
a_{0}^{2}=\frac{2\mathcal{K}f_{R}}{\rho_{0}(1+\omega)(1+f_{T})}.
\end{equation}

\section{Inhomogeneous Scalar Perturbations}

The perturbations play a vital role to convert a difficult
mathematical problem into a simpler one. There are different forms
of perturbations such as isotropic, anisotropic,
homogeneous/inhomogeneous scalar, vector and tensor perturbations.
For successful realization of emergent universe, the EU must display
stable regions against all kinds of perturbations. Several
researchers have adopted these perturbations to examine the stable
state of EU. It is observed that the inhomogeneous perturbations
lead to unstable solutions in $f(R)$ framework (Seahra and
B\"{o}hmer 2009). What will happen in $f(R,T)$ scenario? Will the
stable EU solution exist under the influence of inhomogeneous
perturbations? To answer these queries, here we inspect the stable
modes of EU by implementing inhomogeneous linear perturbations. We
consider Newtonian/longitudinal gauge whose perturbed line element
is expressed by (Huang et al. 2015)
\begin{equation}\label{14}
ds^{2}=(1-2\vartheta)a_{0}^{2}d\tau^{2}-(1-2\varphi)a_{0}^{2}
\left(\frac{1}{1-K\chi^{2}}d\chi^{2}+\chi^{2}d\theta^{2}
+\chi^{2}\sin^{2}\theta d\phi^{2}\right),
\end{equation}
where $\vartheta$ exhibits the Bardeen potential and $\varphi$
represents the perturbation to spatial curvature. The linear
perturbations in matter components yield
\begin{equation}\nonumber
p=p_{0}(1+\delta p),\quad \rho=\rho_{0}(1+\delta\rho),
\end{equation}
where $\delta\rho$ and $\delta p$ depict the perturbed matter
density and pressure, respectively. The harmonic decomposition of
inhomogeneous linear perturbations are (Seahra and B\"{o}hmer 2009)
\begin{eqnarray}\nonumber
\delta p&=&\delta p_{l}(\tau)\Omega_{l}(\mu^{i}),\quad \delta
\rho=\delta \rho_{l}(\tau)\Omega_{l}(\mu^{i}),\\\nonumber
\vartheta&=&\vartheta_{l}(\tau)\Omega_{l}(\mu^{i}),\quad
\varphi=\varphi_{l}(\tau)\Omega_{l}(\mu^{i}).
\end{eqnarray}
Here, $\mu^{i}$ demonstrates the spatial entities
$(\chi,\theta,\phi)$, when summation on $l$ is considered. The
harmonic function $\Omega_{l}(\mu^{i})\equiv\Omega_{l}$ for
different cosmic models describes the following relations
\begin{eqnarray}\nonumber
\Delta\Omega_{l}\equiv-\hbar^{2}\Omega_{l}=\left\{\begin{array}{lll}
-(l^2+1)\Omega_{l},&\quad l^{2}\geq0,&\quad \mathcal{K}=-1,\\
-l^2\Omega_{l},&\quad l^{2}\geq0,&\quad
\mathcal{K}=0,\\-l(l+2)\Omega_{l},&\quad l=0,1,2,...,&\quad
\mathcal{K}=1,\end{array}\right.
\end{eqnarray}
where $\Delta$ acts as the three-dimensional Laplacian operator.

These inhomogeneous perturbations yield discrete spectrum for open
geometry of cosmos while a continuous spectrum is produced for
closed and flat cosmic models (Huang et al. 2015). It is noted that
for $l=0$, one can recover the homogeneous scalar perturbations.
Implementing Taylor series on $f(R,T)$ function and applying
inhomogeneous perturbations, we obtain $\delta R$ and $\delta T$
\begin{eqnarray}\label{15}
&&\delta R=-\frac{2}{a_{0}^{2}}\left(3\ddot{\varphi}
-6\mathcal{K}\varphi-2a_{0}^{2}\hbar^{2}\varphi+a_{0}^{2}\hbar^{2}
\vartheta\right),~\delta T=(1-3\omega)\rho_{0}\delta\rho.
\end{eqnarray}
Substituting inhomogeneous perturbations and Eq.(\ref{15}) in
(\ref{3}), the linearized $\tau\tau$ and diagonal entities
associated with perturbed line element (\ref{14}) provide
\begin{eqnarray}\nonumber
&&(6\mathcal{K}+2a_{0}^{2}\hbar^{2})\varphi f_{R}(R_{0},T_{0})
+a_{0}^{2}\rho_{0}\Big[1-\frac{(\omega-3)}{2}f_{T}(R_{0},T_{0})
+(1-3\omega)\rho_{0}\Big.\\\label{16}&\times&\Big.(1+\omega)
f_{TT}(R_{0},T_{0})\Big]\delta\rho+a_{0}^{2}\hbar^{2}
f_{RR}(R_{0},T_{0})\delta R=0,\\\nonumber
&&2\Big[6\mathcal{K}\varphi-a_{0}^{2}\hbar^{2}(\vartheta-2\varphi)
-3\ddot{\varphi}\Big]f_{R}(R_{0},T_{0})+a_{0}^{2}\rho_{0}
\Big[(1-3\omega)+(3-5\omega)\Big.\\\nonumber&\times&\Big.f_{T}
(R_{0},T_{0})+(\omega+1)(1-3\omega)\rho_{0}f_{TT}(R_{0},T_{0})
+\Big(\frac{9\mathcal{K}}{a_{0}^{2}}+2\hbar^{2}\Big)(1-3\omega)
\Big.\\\nonumber&\times&\Big.f_{RT}(R_{0},T_{0})\Big]\delta\rho
-3a_{0}^{2}\rho_{0}(1-3\omega)f_{RT}(R_{0},T_{0})\delta\ddot{\rho}
-3a_{0}^{2}f_{RR}(R_{0},T_{0})\delta\ddot{R}\\\label{17}&+&3a_{0}^{2}
\Big(\frac{f_{R}(R_{0},T_{0})}{2}+\Big(\frac{2\mathcal{K}}{a_{0}^{2}}
+\hbar^{2}\Big)f_{RR}(R_{0},T_{0})\Big)\delta R=0.
\end{eqnarray}
For perfect matter configuration, the non-diagonal constituents
yield the following connection
\begin{equation}\label{18}
\vartheta(\tau)=\varphi(\tau),
\end{equation}
while anisotropic matter distribution does not satisfy this
relation.

In order to observe $f(R,T)$ theory as a feasible gravitational
theory, one must consider an effective and viable expression of
$f(R,T)$ function. The models of this gravity are displayed in the
following ways (Harko et al. 2011: Harko and Lobo 2019).
\begin{itemize}
\item $f(R,T)=f_{1}(R)+f_{2}(T)$. This choice of $f(R,T)$ function
corresponds to the minimal interaction and can be regarded as a
linear extension to $f(R)$ theory. By adopting any linear
combination of $f_{2}$, various models can be obtained for different
choices of $f_{1}(R)$ function. If we consider $f_{1}(R)=R$ and
$f_{2}(T)=2h(T)$, then the outcomes of this model show consistency
with $\Lambda$CDM cosmological model.
\item $f(R,T)=f_{1}(R)+f_{2}(T)f_{3}(R)$. This form describes
the non-minimal coupling between matter and geometry. The results
obtained from this choice may be different from the minimal coupled
models.
\end{itemize}
To evaluate the stable modes of EU, we adopt the first form of
$f(R,T)$ gravity. The field equations (\ref{16}) and (\ref{17})
corresponding to minimally coupled $f(R,T)$ function yield
\begin{eqnarray}\nonumber
&&(6\mathcal{K}+2a_{0}^{2}\hbar^{2})\varphi f_{1}'(R_{0})
+a_{0}^{2}\rho_{0}\Big[1-\frac{(\omega-3)}{2}f_{2}'(T_{0})
+(1-3\omega)\rho_{0}\Big.\\\label{19}&\times&\Big.(1+\omega)
f_{2}''(T_{0})\Big]\delta\rho+a_{0}^{2}\hbar^{2}
f_{1}''(R_{0})\delta R=0,\\\nonumber
&&2\Big[6\mathcal{K}\varphi-a_{0}^{2}\hbar^{2}(\vartheta-2\varphi)
-3\ddot{\varphi}\Big]f_{1}'(R_{0})+a_{0}^{2}\rho_{0}
\Big[(1-3\omega)+(3-5\omega)\Big.\\\nonumber&\times&\Big.f_{2}'(T_{0})
+(\omega+1)(1-3\omega)\rho_{0}f_{2}''(T_{0})\Big]\delta\rho
-3a_{0}^{2}f_{1}''(R_{0})\delta\ddot{R}+3a_{0}^{2}\\\label{20}&\times&
\Big(\frac{f_{1}'(R_{0})}{2}+\Big(\frac{2\mathcal{K}}{a_{0}^{2}}
+\hbar^{2}\Big)f_{1}''(R_{0})\Big)\delta R=0,
\end{eqnarray}
where $f_{1}'(R)=df_{1}(R)/dR$ and $f_{2}'(T)=df_{2}(T)/dT$.
Inserting Eq.(\ref{15}) in the elimination of $\vartheta$ and
$\delta\rho$ from Eqs.(\ref{19}) and (\ref{20}), it follows that
\begin{eqnarray}\nonumber
&&18\Big[1-\frac{\omega-3}{2}f_{2}'(T_{0})+\rho_{0}(\omega+1)
(1-3\omega)f_{2}''(T_{0})\Big]f_{1}''(R_{0})\varphi^{(iv)}
+\Big[f_{1}''(R_{0})\\\nonumber&\times&\Big\{-6\Big(\hbar^{2}(2
+3\omega+a_{0}^{2})+6\mathcal{K}\Big(1+\frac{1}{a_{0}^{2}}\Big)\Big)
-3\Big(\hbar^{2}(3+7\omega-(\omega-3)a_{0}^{2})\\\nonumber&-&6\mathcal{K}
(\omega-3)\Big(1+\frac{1}{a_{0}^{2}}\Big)\Big)f_{2}'(T_{0})-6\rho_{0}
(1-3\omega)(1+\omega)\Big(6\mathcal{K}\Big(1+\frac{1}{a_{0}^{2}}\Big)
+\hbar^{2}\\\nonumber&\times&(2+a_{0}^{2})\Big)f_{2}''(T_{0})\Big\}
+9f_{1}'(R_{0})\Big\{-1+\frac{\omega-3}{2}f_{2}'(T_{0})-\rho_{0}(1
+\omega)(1-3\omega)\\\nonumber&\times&f_{2}''(T_{0})\Big\}\Big]
\ddot{\varphi}+\Big[f_{1}''(R_{0})\Big\{12\mathcal{K}\hbar^{2}(5
-3\omega)+\frac{72\mathcal{K}^{2}}{a_{0}^{2}}+2a_{0}^{2}\hbar^{2}
(4-3\omega)\\\nonumber&+&f_{2}'(T_{0})\Big(12\mathcal{K}\hbar^{2}
(9-7\omega)+a_{0}^{2}\hbar^{4}(15-13\omega)-\frac{36\mathcal{K}^{2}
(\omega-3)}{a_{0}^{2}}\Big)+f_{2}''(T_{0})\\\nonumber&\times&\Big(4
\hbar^{2}(15\mathcal{K}+2a_{0}^{2}\hbar^{2})+\frac{72\mathcal{K}^{2}}
{a_{0}^{2}}\Big)\Big\}+f_{1}'(R_{0})\Big\{18\mathcal{K}(2-\omega)
+a_{0}^{2}\hbar^{2}(7-6\omega)\\\nonumber&+&f_{2}'(T_{0})\Big(36
\mathcal{K}(1-\omega)+9(3-\omega)+\frac{a_{0}^{2}\hbar^{2}}{2}(27
-25\omega)\Big)+7a_{0}^{2}\hbar^{2}\rho_{0}(1-3\omega)
\\\label{21}&\times&(1+\omega)f_{2}''(T_{0})\Big\}\Big]\varphi=0.
\end{eqnarray}
Substituting the expression of $a_{0}^{2}$ from Eq.(\ref{13}) in
(\ref{21}), the resulting fourth-ordered perturbed equation in terms
of $\varphi$ takes the form
\begin{eqnarray}\nonumber
&&36\mathcal{K}\rho_{0}(1+\omega)\big(1+f_{2}'(T_{0})\big)
\Big[(\omega+1)(1-3\omega)\rho_{0}f_{2}''(T_{0})-\frac{\omega-3}{2}
f_{2}'(T_{0})+1\Big]\\\nonumber&\times&f_{1}'(R_{0})f_{1}''(R_{0})
\varphi^{(iv)}+\Big[f_{1}''(R_{0})\Big\{-6\Big(2\mathcal{K}\hbar^{2}
f_{1}'(R_{0})\big(\rho_{0}(1+\omega)(2+3\omega)(1\\\nonumber&+&f_{2}'(T_{0})
\big)+2\mathcal{K}f_{1}'(R_{0})\big)+6\mathcal{K}\rho_{0}(1+\omega)
\big(1+f_{2}'(T_{0})\big)\big(2\mathcal{K}f_{1}'(R_{0})+\rho_{0}
(1+\omega)\\\nonumber&\times&\big(1+f_{2}'(T_{0})\big)\big)\Big)-3
f_{2}'(T_{0})\Big(2\mathcal{K}\hbar^{2}f_{1}'(R_{0})\big(\rho_{0}(3
+7\omega)(1+\omega)\big(1+f_{2}'(T_{0})\big)\\\nonumber&-&2\mathcal{K}
(\omega-3)f_{1}'(R_{0})\big)-6\mathcal{K}\rho_{0}(1+\omega)\big(1
+f_{2}'(T_{0})\big)\big(2\mathcal{K}f_{1}'(R_{0})+\rho_{0}(1+\omega)
\\\nonumber&\times&\big(1+f_{2}'(T_{0})\big)\big)\Big)-12\mathcal{K}
\rho_{0}(\omega+1)(1-3\omega)f_{2}''(T_{0})\Big(2\hbar^{2}f_{1}'(R_{0})
\big(\mathcal{K}f_{1}'(R_{0})+\rho_{0}\\\nonumber&\times&(1+\omega)
\big(1+f_{2}'(T_{0})\big)\big)+3\rho_{0}(\omega+1)\big(1+f_{2}'(T_{0})\big)
\big(2\mathcal{K}f_{1}'(R_{0})+\rho_{0}(1+\omega)\\\nonumber&\times&\big(1
+f_{2}'(T_{0})\big)\big)\Big)\Big\}+36\mathcal{K}\rho_{0}(1+\omega)
\big(1+f_{2}'(T_{0})\big)f_{1}'(R_{0})^{2}\Big\{\frac{\omega-3}{2}
f_{2}'(T_{0})-1\\\nonumber&-&\rho_{0}(1-3\omega)(1+\omega)f_{2}''(T_{0})
\Big\}\Big]\ddot{\varphi}+\Big[f_{1}''(R_{0})\Big\{24\mathcal{K}^{2}\rho_{0}
(1+\omega)\big(1+f_{2}'(T_{0})\big)\big(\hbar^{2}\\\nonumber&\times&(5
-3\omega)f_{1}'(R_{0})+3\rho_{0}(\omega+1)\big(1+f_{2}'(T_{0})\big)\big)
+8\mathcal{K}^{2}\hbar^{2}(4-3\omega)f_{1}'(R_{0})^{2}\\\nonumber&+&f_{2}'(T_{0})
\Big(24\mathcal{K}^{2}\rho_{0}(\omega+1)\big(1+f_{2}'(T_{0})\big)
\big(\hbar^{2}(9-7\omega)f_{1}'(R_{0})-3\rho_{0}(1+\omega)
(\omega\\\nonumber&-&3)\big(1+f_{2}'(T_{0})\big)\big)+12\mathcal{K}^{2}
\hbar^{4}(5-\omega)f_{1}'(R_{0})^{2}\Big)+8\mathcal{K}^{2}f_{2}''(T_{0})
\Big(\hbar^{2}f_{1}'(R_{0})\big(15\\\nonumber&\times&\rho_{0}(1+\omega)
\big(1+f_{2}'(T_{0})\big)+4\hbar^{2}f_{1}'(R_{0})\big)+9\rho_{0}^{2}
(1+\omega)^{2}\big(1+f_{2}'(T_{0})\big)^{2}\Big)\Big\}
\\\nonumber&+&\mathcal{K}f_{1}'(R_{0})^{2}\Big\{18\mathcal{K}\rho_{0}^{2}(1
+3\omega)(1+\omega)^{2}\big(1+f_{2}'(T_{0})\big)+2\mathcal{K}\hbar^{2}
(7-6\omega)f_{1}'(R_{0})\\\nonumber&+&f_{2}'(T_{0})\Big(\rho_{0}(1
+\omega)\big(1+f_{2}'(T_{0})\big)\big(36\mathcal{K}(1-\omega)+9(3
-\omega)\big)+\mathcal{K}\hbar^{2}(27-25\\\nonumber&\times&\omega)
f_{1}'(R_{0})\Big)+14\mathcal{K}\hbar^{2}\rho_{0}(1+\omega)(1-3\omega)
(7-6\omega)f_{1}'(R_{0})f_{2}''(T_{0})\Big\}\Big]\varphi=0.\\\label{22}
\end{eqnarray}
In the following section, we analyze the existence as well as
stability of Einstein static solutions corresponding to the specific
choices of $f_{1}(R)$ and $f_{2}(T)$ functions.

\section{Stability Analysis of Einstein Universe}

In this section, we evaluate three classes of solutions that can be
regarded as EU models with respect to the conserved and
non-conserved forms of EMT for $f_{1}(R)=R$. First, we consider the
conserved case to obtain the particular form of $f_{2}(T)$ and
investigate the stability of Einstein solution through graphical
analysis. Second, we assume the non-conserved case in which two
different forms of $f_{2}(T)$ are used to examine the stable modes
of EU.
\subsection*{Case I: Conserved EMT}

The modified theories comprising curvature-matter coupling do not
satisfy the conservation law. The continuity equation in the context
of generic FLRW universe model is demonstrated as
\begin{eqnarray}\nonumber
\dot{\rho}+\frac{3(1+\omega)\dot{a}}{a}\rho&=&\frac{-1}{1+f_{T}(R,T)}
\Big[\rho(1+\omega)\dot{f_{T}}(R,T)+\frac{1-\omega}{2}
\dot{\rho}f_{T}(R,T)\Big].\\\label{23}
\end{eqnarray}
Here, we consider that the conservation law holds in $f(R,T)$
gravity and consequently, the differential equation for minimally
coupled $f(R,T)$ model leads to
\begin{equation}\nonumber
(1-\omega)f_{2}'(T)+2(1+\omega)Tf_{2}''(T)=0,
\end{equation}
whose solution yields a unique expression of $f_{2}(T)$ for which
EMT remains conserved and it is given by
\begin{equation}\label{24}
f_{2}(T)=\Big(\frac{1+\omega}{1+3\omega}\Big)T^{\frac{1+3\omega}{2
(1+\omega)}}c_{1}+c_{2}=0,
\end{equation}
with $c_{1}$ and $c_{2}$ as integration constants. Implementing this
solution in Eq.(\ref{22}) with $f_{1}(R)=R$, we obtain the
inhomogeneous perturbed differential equation of the form
\begin{equation}\label{25}
\mathcal{A}_{1}\varphi-\mathcal{A}_{2}\ddot{\varphi}=0,
\end{equation}
where $\mathcal{A}_{1}$ and $\mathcal{A}_{2}$ are
\begin{eqnarray}\nonumber
\mathcal{A}_{1}&=&\mathcal{K}\Big[18\mathcal{K}\rho_{0}^{2}(1+3\omega)
(1+\omega)^{2}\Big\{1+\frac{c_{1}}{2}\Big((1-3\omega)\rho_{0}\Big)^{\frac{-1
+\omega}{2+2\omega}}\Big\}+2\mathcal{K}\hbar^{2}(7\\\nonumber&-&6\omega)
+\frac{c_{1}}{2}\Big((1-3\omega)\rho_{0}\Big)^{\frac{-1+\omega}{2+2\omega}}
\Big\{9\rho_{0}(1+\omega)\big(4\mathcal{K}(1-\omega)+3-\omega\big)\Big(1
+\frac{c_{1}}{2}\\\nonumber&\times&\Big((1-3\omega)\rho_{0}\Big)^{\frac{-1
+\omega}{2+2\omega}}\Big)+\mathcal{K}\hbar^{2}(27-25\omega)\Big\}+\frac{7
\mathcal{K}\hbar^{2}(\omega-1)(7-6\omega)c_{1}}{2}\\\nonumber&\times&
\Big((1-3\omega)\rho_{0}\Big)^{\frac{-1+\omega}{2+2\omega}}\Big],
\\\nonumber\mathcal{A}_{2}&=&36\mathcal{K}\rho_{0}(\omega+1)\Big[1
+\frac{c_{1}}{2}\Big(\rho_{0}(1-3\omega)\Big)^{\frac{-1
+\omega}{2+2\omega}}\Big]^{2}.
\end{eqnarray}
The solution of Eq.(\ref{25}) is
\begin{equation}\nonumber
\varphi(\tau)=a_{1}e^{\varpi_{1}\tau}+a_{2}e^{-\varpi_{1}\tau},
\end{equation}
where $a_{1}$ and $a_{2}$ are constants of integration. The
parameter $\varpi$ manifests the frequency of perturbation
represented by
\begin{equation}\label{26}
\varpi_{1}^{2}=\mathcal{A}_{1}/\mathcal{A}_{2}.
\end{equation}

The existence of unstable/stable eras of EU is based only on the
exponential growth of perturbations. The inequality
$\varpi_{1}^{2}>0$ leads to the unstable solutions while the stable
ones exist for $\varpi_{1}^{2}<0$. When $l=0$, the frequency
corresponding to the homogenous perturbations is obtained and in
general relativistic limit, i.e., $c_{1}=0$, this frequency reduces
to
\begin{equation}\nonumber
\varpi_{1}^{2}=\frac{\mathcal{K}}{2}\rho_{0}(1+\omega)(1+3\omega),
\end{equation}
which yields the stable solutions for $-1<\omega<-\frac{1}{3}$
(B\"{o}hmer and Lobo 2009). To inspect the stable eras of EU
graphically, we choose current value of $\rho_{0}$ as $\rho_{0}=0.3$
(Ade et al. 2016). Figure \textbf{1} demonstrates the existence of
stable EU modes for both closed and open universe models with
respect to two different values of $l$. It is observed that the
stability of EU increases towards positive values of $\omega$ with
the increasing value of $l$ in the framework of closed cosmic model
while it slightly reduces for $\mathcal{K}=-1$ as $l^{2}$ enhances.
It is also found that more stable modes exist for positive and
negative values of $c_{1}$ with respect to $\mathcal{K}=1$ and
$\mathcal{K}=-1$, respectively. However, in both plots, the stable
EU regions appear for $\omega>-1$ which is consistent with GR.
\begin{figure}\center
\epsfig{file=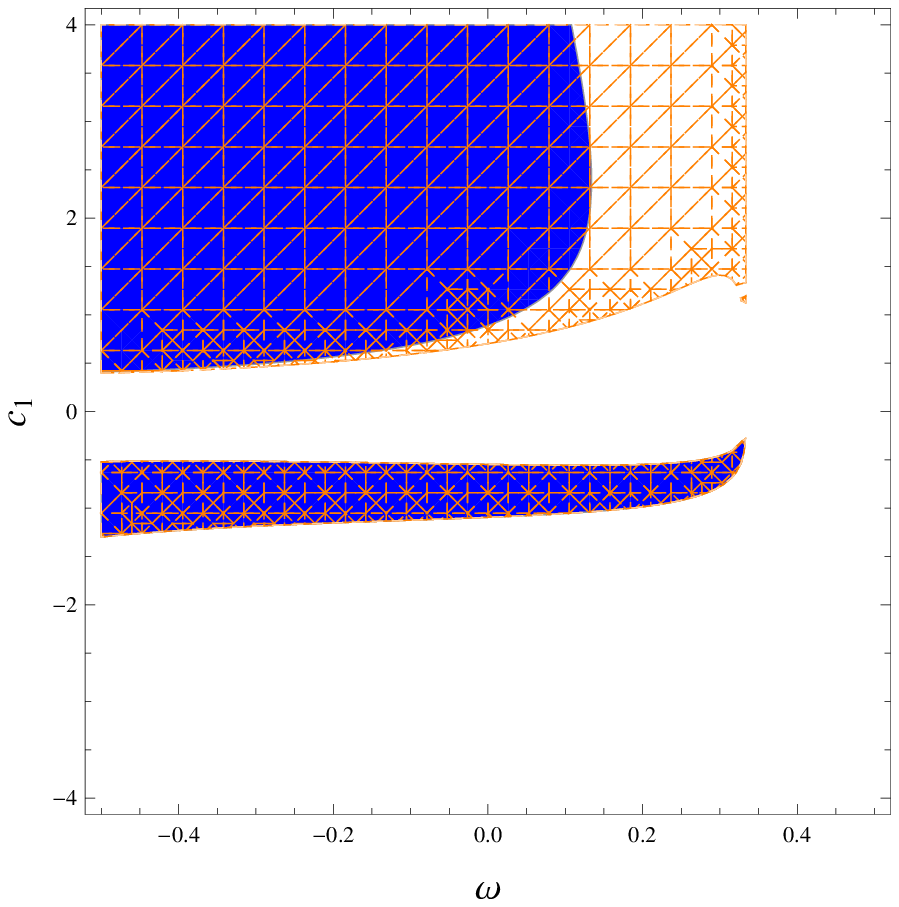,width=0.48\linewidth}
\epsfig{file=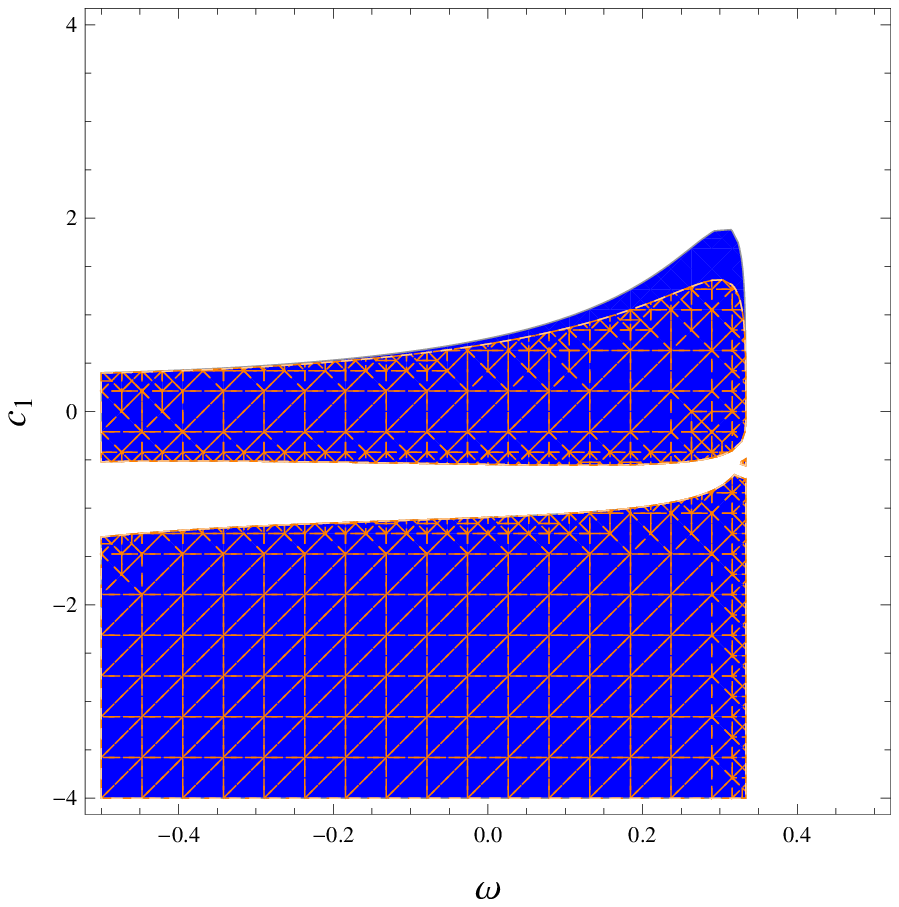,width=0.48\linewidth}\\
\caption{Stability of EU in $(\omega,c_{1})$ space with $l=2$
(blue), $l=15$ (orange) for $\mathcal{K}=1$ (left) and $l^{2}=2$
(blue), $l^{2}=15$ (orange) for $\mathcal{K}=-1$ (right).}
\end{figure}

\subsection*{Case II: Non-conserved EMT}

Here, we inspect the stable Einstein static solution when covariant
divergence of EMT is not zero. For this purpose, we assume two
particular choices of $f(R,T)$ function that describe a direct
relation between $R$ and $T$. First, we take
\begin{equation}\label{27}
f(R,T)=R+m\sqrt{T},
\end{equation}
where $m$ is a coupling constant. Shabani and Ziaie (2017) have used
this model to examine the solution for homogeneous perturbations and
determined that stable modes of EU are obtained only for positive
values of $m$. Inserting this model in Eq.(\ref{22}), the equation
in $\varphi$ is acquired whose solution provides the frequency as
follows
\begin{equation}\label{28}
\varpi_{2}^{2}=\mathcal{A}_{3}/\mathcal{A}_{4},
\end{equation}
where $\mathcal{A}_{i}$'s $(i=3,4)$ are represented by
\begin{eqnarray}\nonumber
\mathcal{A}_{3}&=&\mathcal{K}\Big[18\mathcal{K}\rho_{0}^{2}
(1+3\omega)(1+\omega)^{2}\Big(1+\frac{m}{2\sqrt{\rho_{0}(1
-3\omega)}}\Big)+2\mathcal{K}\hbar^{2}(7-6\omega)
\\\nonumber&+&\frac{m}{2\sqrt{\rho_{0}(1
-3\omega)}}\Big\{9\rho_{0}(1+\omega)\big(4
\mathcal{K}(1-\omega)+3-\omega\big)\Big(1+\frac{m}{2\sqrt{\rho_{0}(1
-3\omega)}}\Big)\\\nonumber&+&\mathcal{K}\hbar^{2}(27-25
\omega)\Big\}-\frac{7\mathcal{K}\hbar^{2}(\omega+1)(7-6
\omega)m}{2\sqrt{\rho_{0}(1-3\omega)}}\Big],
\\\nonumber\mathcal{A}_{4}&=&-36\mathcal{K}\rho_{0}(1+
\omega)\Big(1+\frac{m}{2\sqrt{\rho_{0}(1
-3\omega)}}\Big)\Big(\frac{m(\omega-1)}{2\sqrt{\rho_{0}(1
-3\omega)}}-1\Big).
\end{eqnarray}

The graphical interpretation of stable modes in non-conserved state
for different values of $l$ is exhibited in Figure \textbf{2}. For
closed universe model, it is observed that stability increases as
$l$ increases for positive values of $m$ whereas it almost remains
the same in case of open cosmic model. However, for
$\mathcal{K}=-1$, more stable eras exist in comparison with
$\mathcal{K}=1$. For $m=0=l$, the frequency can retrieve the results
of GR as displayed in the case of conserved EMT.
\begin{figure}\center
\epsfig{file=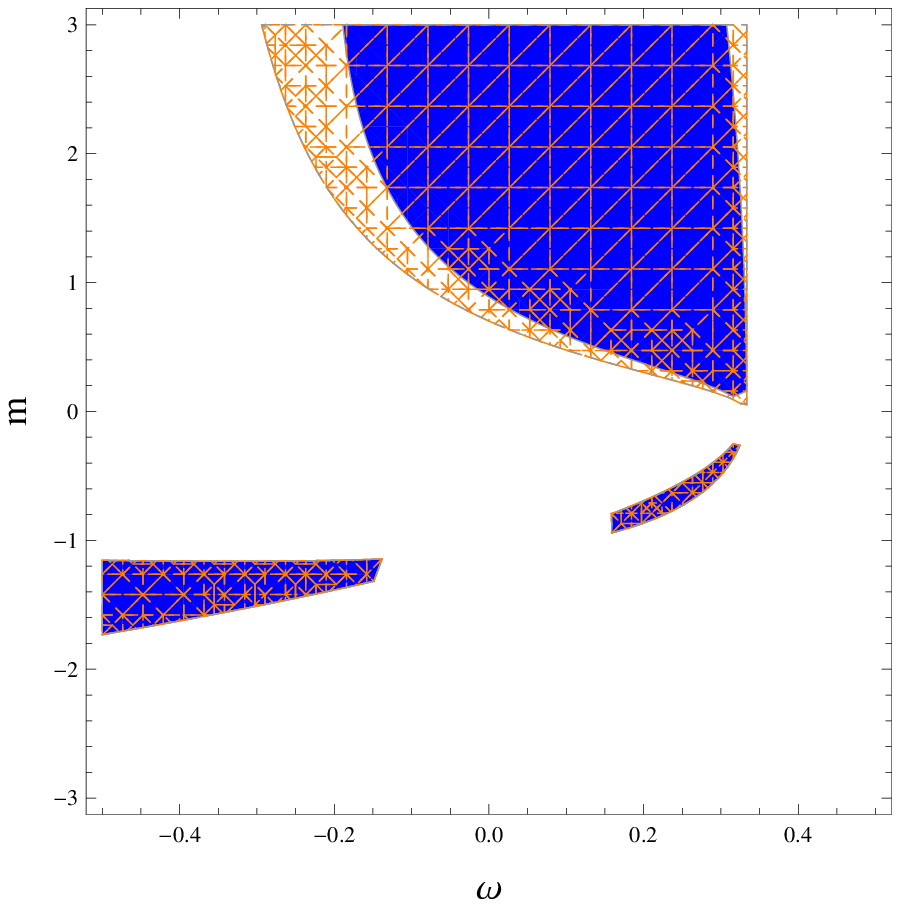,width=0.47\linewidth}
\epsfig{file=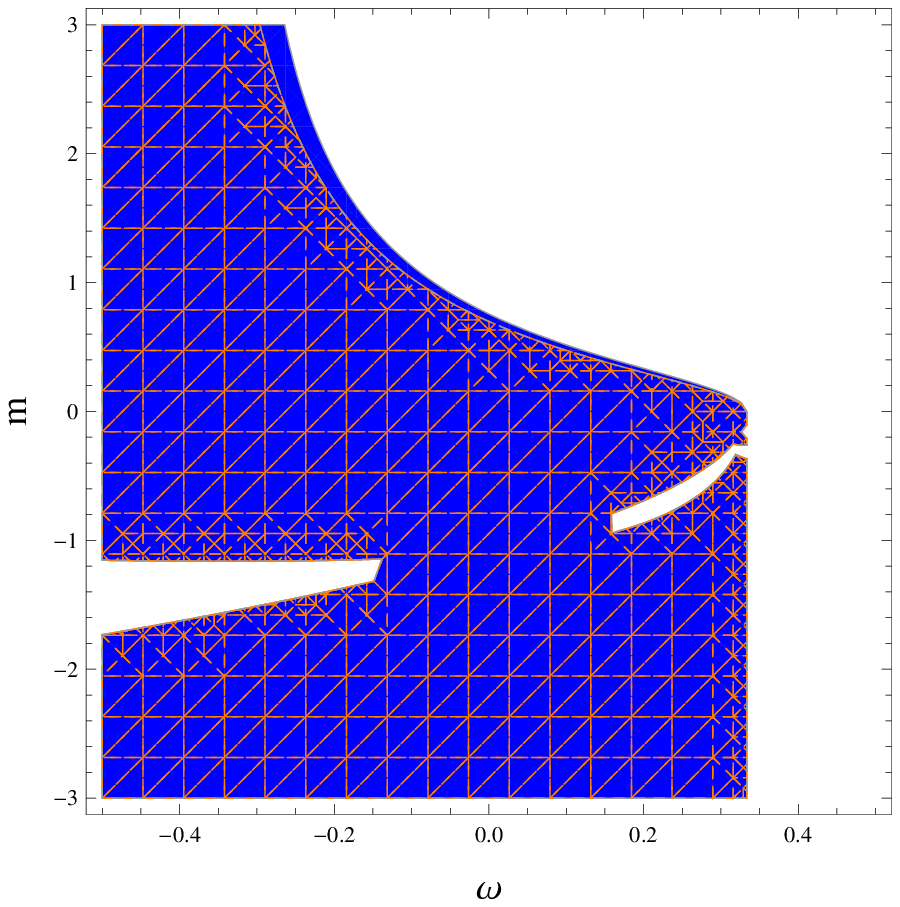,width=0.47\linewidth}\\
\caption{Stability of EU in $(\omega,m)$ space with $l=2$ (blue),
$l=15$ (orange) for $\mathcal{K}=1$ (left) and $l^{2}=2$ (blue),
$l^{2}=15$ (orange) for $\mathcal{K}=-1$ (right).}
\end{figure}

Now, we consider power-law model of $f(R,T)$ gravity presented by
(Shabani and Ziaie 2017)
\begin{equation}\label{29}
f(R,T)=R+\alpha T^{\beta},
\end{equation}
with arbitrary constants displayed by $\alpha$ and $\beta$. For this
model, the solution of differential equation (\ref{22}) yields the
following form of frequency
\begin{equation}\label{30}
\varpi_{3}^{2}=\mathcal{A}_{5}/\mathcal{A}_{6},
\end{equation}
where $\mathcal{A}_{5}$ and $\mathcal{A}_{6}$ are
\begin{eqnarray}\nonumber
\mathcal{A}_{5}&=&\mathcal{K}\Big[18\mathcal{K}\rho_{0}^{2}
(1+\omega)^{2}(1+3\omega)\Big(1+\alpha\beta\big((1
-3\omega)\rho_{0}\big)^{\beta-1}\Big)+2\mathcal{K}\hbar^{2}(7-6\omega)
\\\nonumber&+&\alpha\beta\big((1-3\omega)\rho_{0}\big)^{\beta-1}
\Big\{9\rho_{0}(1+\omega)\Big(4\mathcal{K}(1-\omega)+3-\omega\Big)
\Big(1+\alpha\beta\big(\rho_{0}(1\\\nonumber&-&3\omega)\big)^{\beta
-1}\Big)+\mathcal{K}\hbar^{2}(27-25\omega)\Big\}+14\mathcal{K}
\hbar^{2}(1+\omega)(7-6\omega)(\beta-1)\\\nonumber&\times&\big(\rho_{0}
(1-3\omega)\big)^{\beta-1}\Big],
\\\nonumber\mathcal{A}_{6}&=&-36\mathcal{K}\rho_{0}(1+\omega)\Big(1
+\alpha\beta\big(\rho_{0}(1-3\omega)\big)^{\beta-1}\Big)\Big[\alpha
\beta\Big(\frac{\omega-3}{2}-(1+\omega)(\beta\\\nonumber&-&1)\Big)
\Big(\rho_{0}(1-3\omega)\Big)^{\beta-1}-1\Big].
\end{eqnarray}
For $\alpha=0$, this frequency reduces to GR.
\begin{figure}\center
\epsfig{file=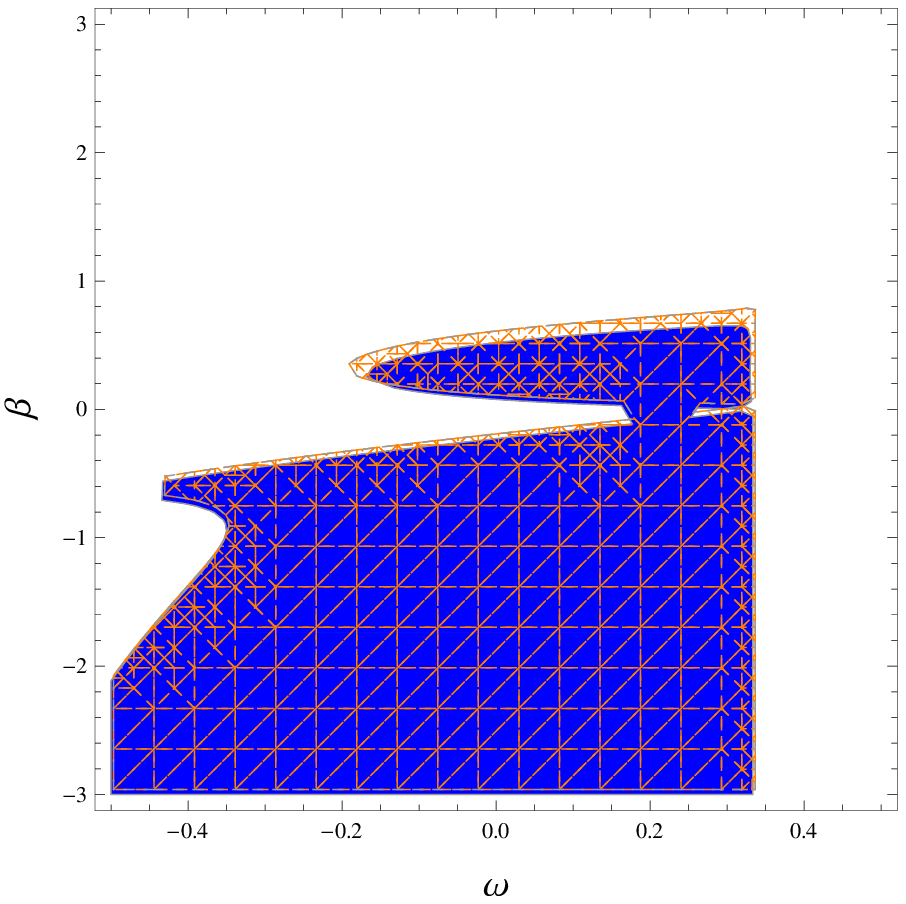,width=0.46\linewidth}
\epsfig{file=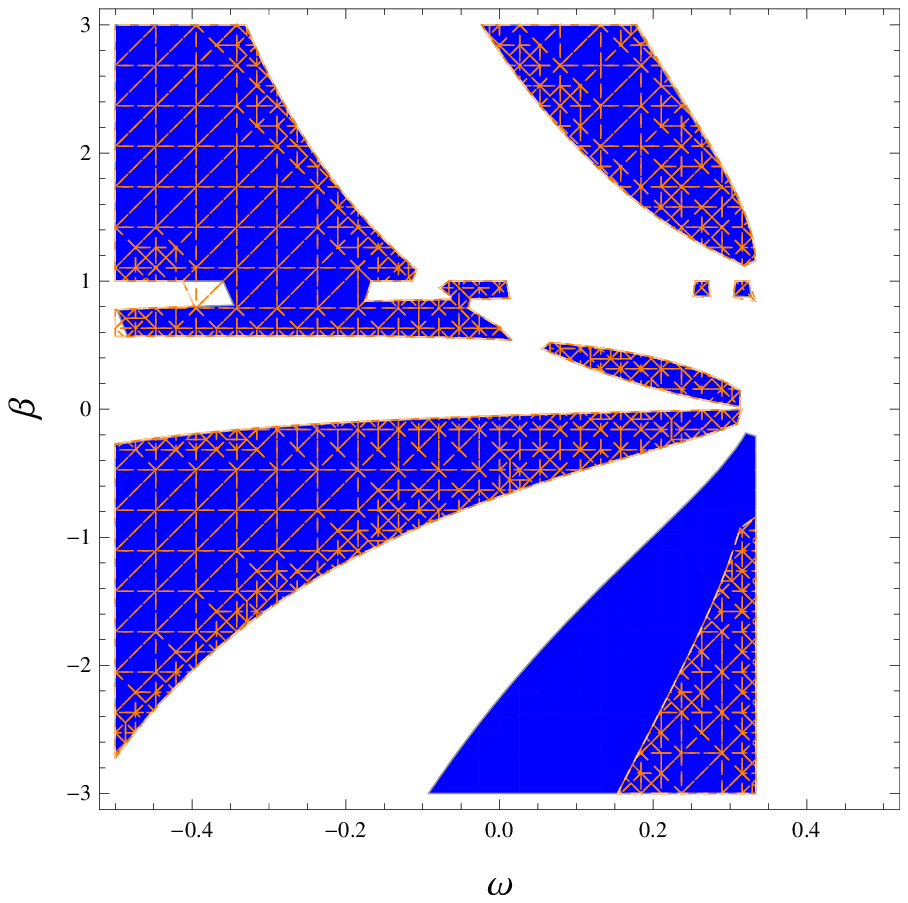,width=0.46\linewidth}\\
\caption{Stability of EU in $(\omega,\beta)$ space for
$\mathcal{K}=1$ with $\alpha=1$ (left), $\alpha=-1$ (right), $l=2$
(blue) and $l=15$ (orange).}
\end{figure}
\begin{figure}\center
\epsfig{file=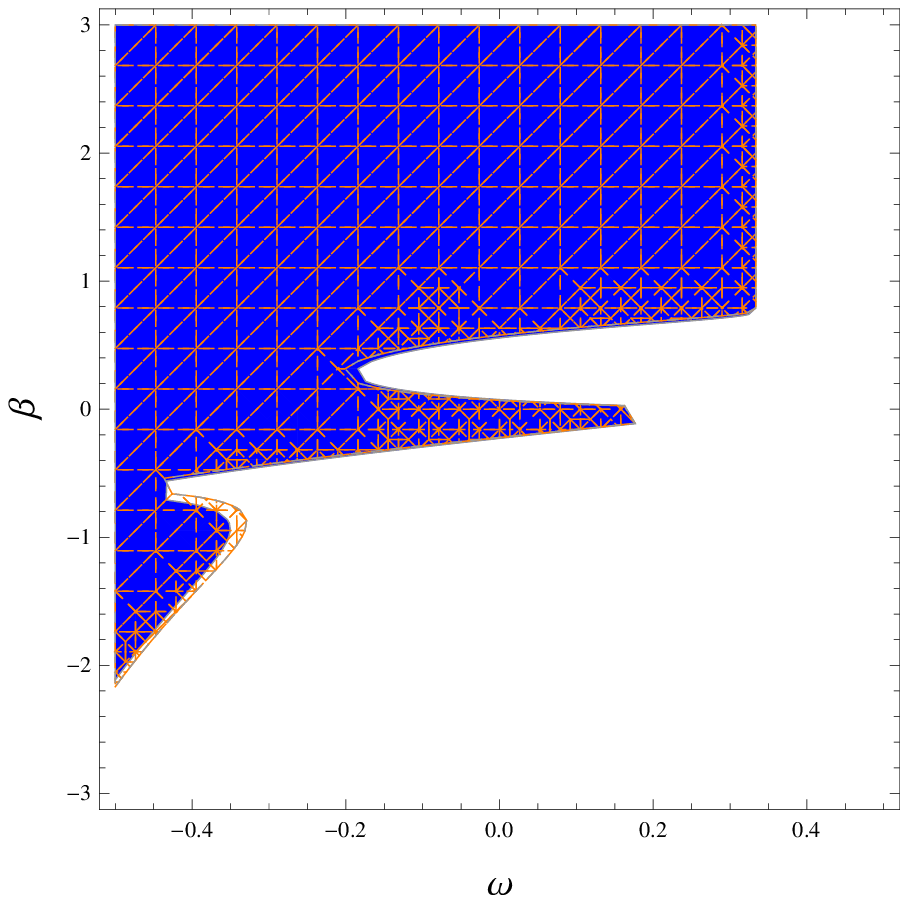,width=0.46\linewidth}
\epsfig{file=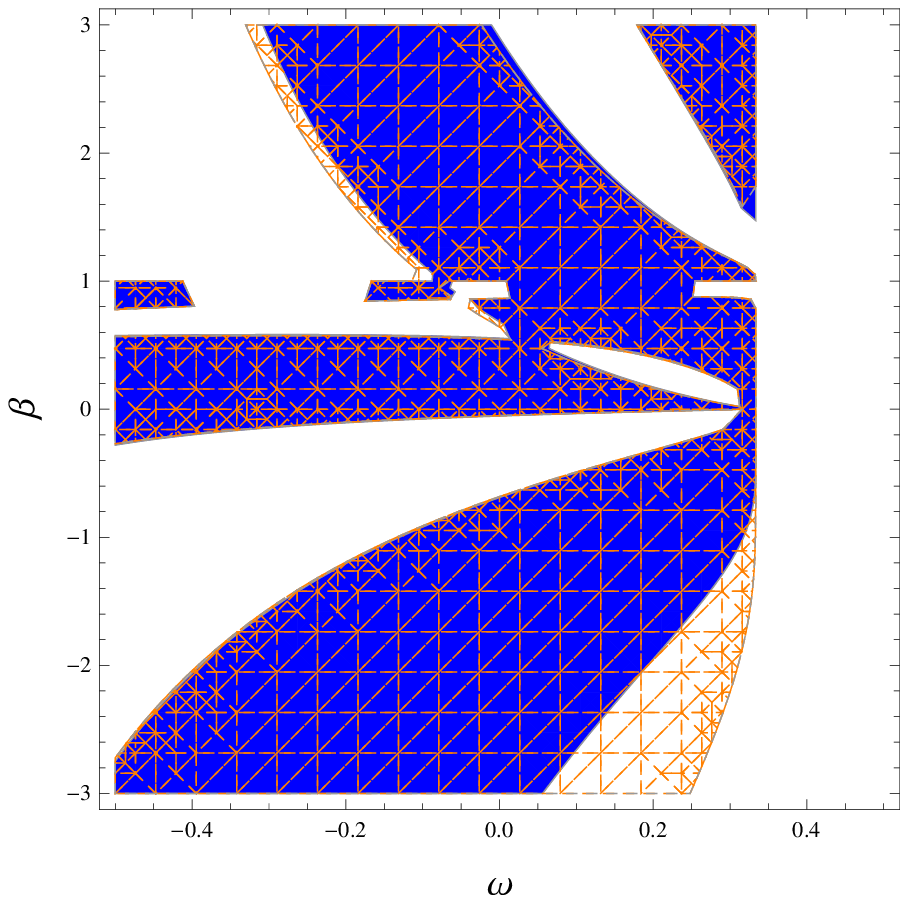,width=0.46\linewidth}\\
\caption{Stability of EU in $(\omega,\beta)$ space for
$\mathcal{K}=-1$ with $\alpha=1$ (left), $\alpha=-1$ (right),
$l^{2}=2$ (blue) and $l^{2}=15$ (orange).}
\end{figure}

Figures \textbf{3} and \textbf{4} manifest the stable eras of EU for
distinct values of $l$ and $\alpha$ with respect to closed and open
geometries of cosmos, respectively. From these Figures, we observe
that the stability slightly enhances and reduces with increasing
values of $l$ in case of closed cosmic model for positive and
negative value of $\alpha$, respectively. For $\mathcal{K}=-1$, both
values of $\alpha$ correspond to increasing stable modes of EU as
$l$ increases. It is also found that more stable regions appear for
positive and negative values of $\beta$ in the background of
$\mathcal{K}=1$ and $\mathcal{K}=-1$, respectively. From these
graphical analysis, we can conclude that the considered $f(R,T)$
models provide more stable regions of EU with inhomogeneous
perturbations as compared to homogeneous perturbations (Shabani and
Ziaie 2017).

\section{Concluding Remarks}

The conjecture of emergent universe has been identified as a
feasible alternative to the big-bang singularity and modified
theories have been proved successful tool to discuss this
conjecture. In this paper, we have analyzed the existence of stable
EU in the domain of different $f(R,T)$ models. The static EU
solutions have been examined by employing scalar inhomogeneous
perturbations characterized by linear EoS. We have obtained the
second order perturbed differential equations for three specific
$f(R,T)$ functions with respect to conservation and non-conservation
of EMT.

For the conserved EMT, we have evaluated a peculiar expression of
$f(T)$ for which the continuity equation satisfies in $f(R,T)$
framework. We have examined the EU solutions against integration
constant $c_{1}$ for different values of $l$. We have observed that
more stable eras of EU appear for positive and negative values of
$c_{1}$ with respect to closed and open FLRW models, respectively.
In case of non-conserved EMT, we have considered the power-law forms
of $f(R,T)$ gravity and derived the stable EU solutions for
appropriate choices of model parameters. It is worthy to mention
that our solutions can be transformed to homogeneous perturbations
for $l=0$ and to GR when the model parameters become zero.

Shabani and Ziaie (2017) investigated the presence of stable EU
regions in the same gravity using dynamical system approach and
homogeneous linear perturbations. They found that in comparison with
$f(R)$ gravity in which generally unstable solutions appear (Seahra
and B\"{o}hmer 2009), some particular $f(R,T)$ models lead to stable
EU regions. From our graphical analysis, we conclude that using
scalar inhomogeneous perturbations, more stable regions exist in
$f(R,T)$ gravity as compared to homogeneous linear perturbations
(Shabani and Ziaie 2017). It is noticed that our all stable modes of
EU lie in the interval $-1<\omega<0.35$ which is consistent with GR.
Hence, $f(R,T)$ gravity can provide such environment in which EU is
associated with asymptotic emergent universe conjecture. It would be
worthwhile to discuss this scenario on the ground of anisotropic
perturbations in the same gravity.

\vspace{.25cm}

{\bf Acknowledgment}

\vspace{0.25cm}

One (AW) of us would like to thank the Higher Education Commission,
Islamabad, Pakistan for its financial support through the {\it
Indigenous Ph.D. 5000 Fellowship Program Phase-II, Batch-III.}

\end{document}